\documentclass[runningheads]{llncs}
\usepackage[T1]{fontenc}
\usepackage{graphicx}
\usepackage{booktabs}
\usepackage{color}
\begin{document}
\title{MS-MT: Multi-Scale Mean Teacher with Contrastive Unpaired Translation for Cross-Modality Vestibular Schwannoma and Cochlea Segmentation}

\titlerunning{Multi-scale Mean Teacher}
%
%

\author{Ziyuan Zhao\inst{1, 2, 3}
\and Kaixin Xu\inst{1}
\and Huai Zhe Yeo\inst{1,4,}\thanks{This work was
done when Huai Zhe was an intern at I2R, A*STAR.}
\and Xulei Yang\inst{1, 2} 
\and Cuntai Guan\inst{3}}

\authorrunning{Zhao et al.}
%
\institute{Institute for Infocomm Research (I$^2$R), A*STAR, Singapore \and
Artificial Intelligence, Analytics And Informatics (AI$^3$), A*STAR, Singapore \and
Nanyang Technological University, Singapore \and
National University of Singapore, Singapore 
}


%
\maketitle              

\begin{abstract}
Domain shift has been a long-standing issue for medical image segmentation.
Recently, unsupervised domain adaptation (UDA) methods have achieved promising cross-modality segmentation performance by distilling knowledge from a label-rich source domain to a target domain without labels. 
In this work, we propose a multi-scale self-ensembling based UDA framework for automatic segmentation of two key brain structures~\emph{i.e.,} Vestibular Schwannoma (VS) and Cochlea on high-resolution T2 images.
First, a segmentation-enhanced contrastive unpaired image translation module is designed for image-level domain adaptation from source T1 to target T2.
Next, multi-scale deep supervision and consistency regularization are introduced to a mean teacher network for self-ensemble learning to further close the domain gap.
Furthermore, self-training and intensity augmentation techniques are utilized to mitigate label scarcity and boost cross-modality segmentation performance.
Our method demonstrates promising segmentation performance with a mean Dice score of $83.8\%$ and $81.4\%$ and an average asymmetric surface distance (ASSD) of $0.55$ mm and $0.26$ mm for the VS and Cochlea, respectively in the validation phase of the crossMoDA $2022$ challenge.


\keywords{Medical image segmentation \and Unsupervised domain adaptation \and Vestibular Schwannoma}


\end{abstract}
\section{Introduction}
Medical image segmentation plays a vital role in the field of medical image analysis, delivering valuable information for diagnostic analysis and treatment planning~\cite{hesamian2019deep}. 
Accurate segmentation and measurement of Vestibular Schwannoma (VS) and Cochlea from MRI can assist in VS treatment planning, improving clinical workflow~\cite{shapey2019artificial}.
To this end, researchers have turned to deep learning as a solution to perform autonomous segmentation of VS and Cochlea. 
Recent findings also suggest that high-resolution T2 (hrT2) MRI could be a safer and more cost-efficient alternative to contrast-enhanced T1 (ceT1) MRI. 
However, the large domain shift between MRI images with different contrasts coupled with the costly and laborious process of re-annotating medical image scans on another modality, makes it difficult for deep learning to generalize well across both domains.
Therefore, we are encouraged to perform unsupervised domain adaptation~(UDA) and conduct VS and Cochlea segmentation in the hrT2 domain by leveraging labeled ceT1 scans and unlabeled hrT2 scans. 
In this work, we propose an effective and intuitive UDA method based on image translation and self-ensembling learning. 
Firstly, we translate MRI scans from the ceT1 domain to the hrT2 domain using a modified CUT model~\cite{park2020cut} for image-level adaptation. 
Then, a nnU-Net~\cite{nnunet} is trained on synthetic hrT2 images to generate pseudo annotations for unlabeled hrT2 images. 
Finally, we build a multi-scale mean-teacher (MS-MT) network~\cite{meanteacher,li2021hierarchical} to transfer multi-level knowledge from the teacher model to the student model for improving the cross-modality segmentation performance. 
The experimental results show that the proposed UDA network can greatly reduce the domain gap, achieving promising segmentation performance on hrT2 scans.

\section{Related Work}

To bridge the domain gap across different modalities, many UDA methods have been developed in the medical imaging field to align the distributions between modalities from different perspectives, including image-level alignment~\cite{zhu2017unpaired,huo2018adversarial,zhang2018task}, feature-level alignment~\cite{tzeng2014deep,long2015learning,dou2018pnp,wang2022unsupervised} and their combinations~\cite{hoffman2018cycada,chen2020unsupervised}. Image alignment is a practical UDA approach that translates source images to appear as if they were sampled from the target domain or vice versa. CycleGAN~\cite{zhu2017unpaired} is mainly used to achieve unpaired image-to-image translation for UDA in medical image segmentation~\cite{huo2018adversarial,zhang2018task}. For instance, Huo~\emph{et al.}~\cite{huo2018adversarial} leveraged CycleGAN to perform MRI to CT synthesis for enabling CT splenomegaly segmentation without using target labels. Zhang~\emph{et al.}~\cite{zhang2018task} proposed a task-driven generative adversarial network by introducing segmentation consistency into the DRR to X-ray translation process to achieve X-ray segmentation with the pre-trained DRR segmentation models. Recently, contrastive learning has demonstrated a powerful capacity for unsupervised visual representation learning in various computer vision tasks~\cite{jaiswal2021survey}. Park~\emph{et al.}~\cite{park2020cut} first proposed CUT to use contrastive learning for image synthesis, in which, the mutual information between original and generated patches was maximized to keep the content unchanged after translation. Several studies have applied CUT in cross-domain medical image analysis~\cite{choi2022using,liu2022unsupervised}. Therefore, in this study, we use CUT for image-level adaptation. Another line of research in UDA is to align the feature distributions across domains from different aspects, including explicit discrepancy minimization~\cite{tzeng2014deep,long2015learning} and implicit adversarial learning~\cite{ganin2016domain,dou2018pnp}. For example, Dou~\emph{et al.}~\cite{dou2018pnp} proposed to fine-tune specific feature layers via adversarial learning for cross-modality cardiac segmentation. To achieve image and feature adaptation, CyCADA~\cite{hoffman2018cycada} was proposed to bridge the appearance gap using CycleGAN and align the feature spaces with adversarial learning separately. Chen~\emph{et al.}~\cite{chen2020unsupervised} further introduced additional discriminators in CycleGAN to achieve concurrent feature and image adaptation. These UDA methods have been used in cross-modality VS and Cochlea Segmentation for closing the domain gap~\cite{CrossModaChallenge}. More recently, the usage of weak labels, such as scribbles on the target domain, for weakly-supervised domain adaptation~\cite{dorent2020scribble} has also shown much promise, receiving attention from the community.

On the other hand, since the labels on the target domain are not accessible, not-so-supervised learning methods~\cite{tajbakhsh2020embracing}, such as semi-supervised learning (SSL)~\cite{bai2017semi,meanteacher} can be leveraged to relax the dependence on target labels for improving the adaptation performance~\cite{zou2018unsupervised,zhao2021mt,zhao2022le}. Zou~\emph{et al.}~\cite{zou2018unsupervised} built a self-training pseudo-labeling framework for UDA, in which, pseudo-target labels are generated to retrain the model iteratively for improving the UDA performance. Zhao~\emph{et al}~\cite{zhao2021mt,zhao2022le} investigated the source label scarcity problem in UDA, and leveraged self-ensembling models~\cite{meanteacher} to address annotation scarcity on both domains for label-efficient UDA. These works suggest that SSL techniques can also be leveraged in UDA to improve adaptation performance. In this regard, pseudo-labeling and self-ensembling learning methodologies were explored in our UDA framework.

\section{Methods}

\begin{figure}[t]
    \centering
    \includegraphics[width = \columnwidth]{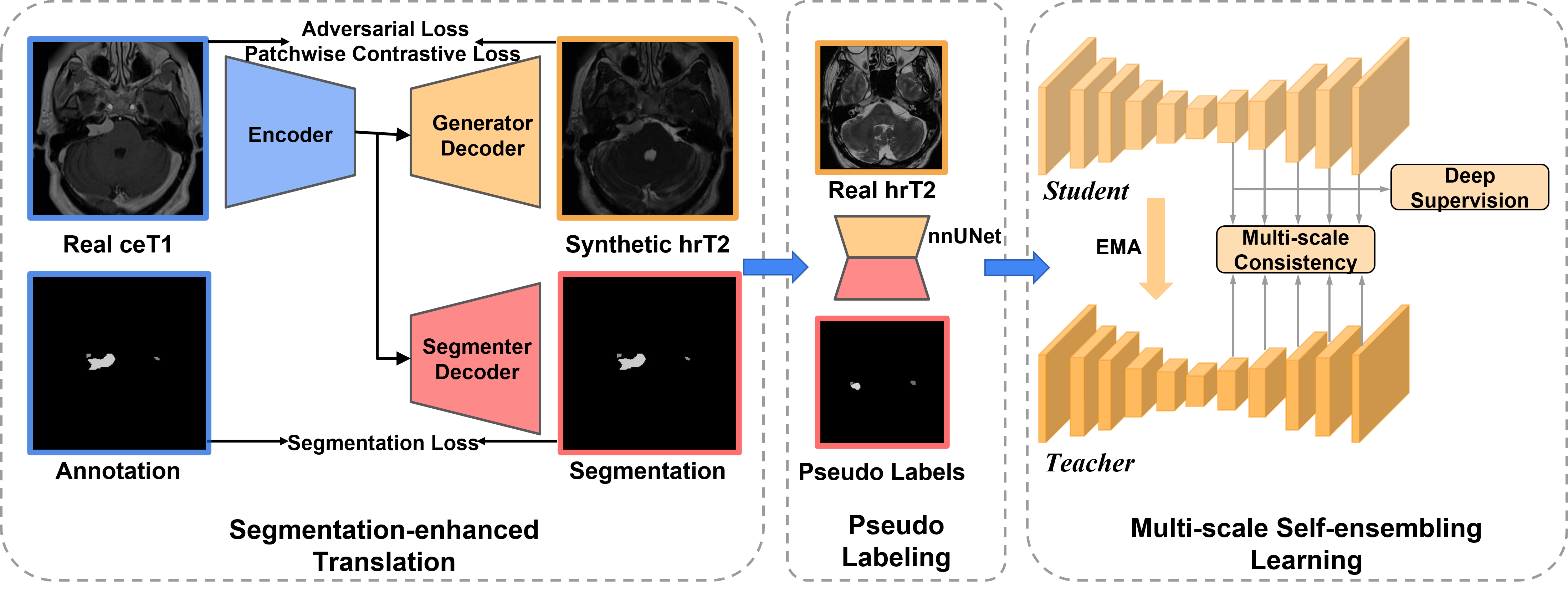}
    \caption{The overview of our proposed method. First, synthetic hrT2 images are generated with the proposed segmentation-enhanced translation network. Then, we employ a nnU-Net for training on synthetic hrT2 images and generate pseudo labels for unlabeled hrT2 images. Finally, a multi-scale mean teacher network is employed to further close the domain gap.}
    \label{fig:pipeline}
\end{figure}

Given an unpaired dataset of two modalities,~\emph{i.e.,} annotated ceT1 MRI images $\mathcal{D}_{s}=\left\{\left(\mathbf{x}_{i}^{s}, y_{i}^{s}\right)\right\}_{i=1}^{N}$ and non-annotated hrT2 MRI scans  $\mathcal{D}_{t}=\left\{\left(\mathbf{x}_{i}^{t}\right)\right\}_{i=1}^{M}$, sharing the same classes (VS and Cochlea), we aim to exploit $\mathcal{D}_{s}$ and $\mathcal{D}_{t}$ for unsupervised domain adaptation to enhance the cross-modality segmentation performance of the VS and Cochlea on hrT2 MRI images. The overview of our UDA framework is shown in Fig.~\ref{fig:pipeline}.




\subsection{Segmentation-enhanced translation}

\begin{figure}[t]
    \centering
    \includegraphics[width =0.9\columnwidth]{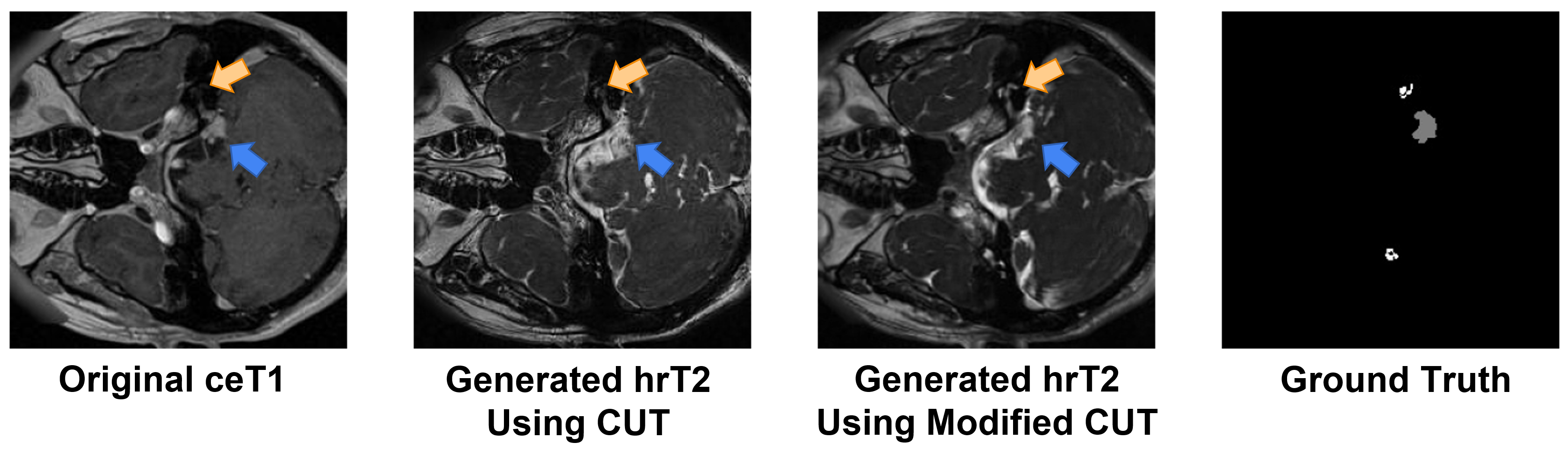}
    \caption{Visual comparison of the synthetic hrT2 images by different methods.}
    \label{fig:generation_comparsion}
\end{figure}

To close the domain gap across the modalities, we first conduct image-level domain adaptation to generate synthetic target samples. 
In this regard, the model trained on the synthetic target images will be used for VS and Cochlea segmentation on real hrT2 scans.
For unpaired image-to-image translation, we adopt the Contrastive Unpaired Translation (CUT)~\cite{park2020cut} as our backbone since it is faster and less memory-intensive than CycleGAN~\cite{zhu2017unpaired}.
Moreover, we enhance the 2D CUT with an additional segmentation decoder for maintaining the structural information of VS and Cochlea (see Fig.~\ref{fig:pipeline}).
Specifically, a ResNet-based generator is used to translate images from the source domain to the target domain, while a PatchGAN discriminator is employed to distinguish between the real and generated images~\cite{park2020cut}. We follow the SIFA architecture~\cite{chen2020unsupervised} and connect two layers of the encoder, specifically at the last layer and the layer before the last downsampling, with the segmenter decoder to generate multi-level segmentation predictions. The segmentation loss can help the encoder focus more on areas related to the segmentation task and preserve the structure of VS and Cochlea in the translated images. In Fig.~\ref{fig:generation_comparsion}, we can observe that the modified CUT can better preserve the shape of the VS and Cochlea in comparison with the original CUT framework.

\subsection{Intensity augmentation and pseudo-labeling}


Considering that tumors exhibit T2 heterogeneous signal intensity~\cite{VS_hypointense} and Cochleas show T2 hyperintense signal intensity~\cite{Cochlea_hyperintense}, we perform intensity augmentation~(IA) and generate augmented data for diversifying the training distributions, thereby improving the model generalizability. 
Using the generated hrT2 images and the ground truth annotations, the signal intensities of both the VS and Cochlea were muted and intensified by $50\%$ respectively, doubling the training data.
On the other hand, to boost the segmentation performance on real hrT2 images, we adopt a Pseudo-Labeling (PL) strategy to leverage unlabeled hrT2 images by generating pseudo hrT2 annotations. 
We employ a 3D full resolution nnU-Net~\cite{nnunet} for training on synthetic hrT2 images and augmented images, which is then used on unlabeled hrT2 images to generate pseudo-labels.

\subsection{Multi-scale self-ensembling learning}
To further utilize all available data, we propose to take advantage of the self-ensembling network,  mean teacher (MT)~\cite{meanteacher}, in which, a teacher model is constructed with the same architecture as the student model, and updated with an exponential moving average (EMA) of the student parameters during training. 
In our training process, the outputs of the student and teacher models with different perturbations are optimized to be consistent by minimizing the difference using the mean square error (MSE) loss.
Inspired by the great success of multi-scale learning in medical image analysis~\cite{dou20173d,li2021hierarchical,zhao2022mmgl}, we follow~\cite{li2021hierarchical} to construct a multi-scale mean teacher (MS-MT) network to leverage multi-scale predictions for deep supervision and consistency regularization. 
We implement 3D full resolution nnU-Net~\cite{nnunet} as the backbone for both teacher and student networks, in which, auxiliary layers are connected to each block of the last five blocks to obtain multi-scale predictions (see Fig.~\ref{fig:pipeline}).

\section{Experiments and Results}

\subsection{Dataset and pre-processing}
The training dataset released by the MICCAI challenge crossMoDA 2022~\cite{CrossModaChallenge,CrossModaData1,CrossModaData2,CrossModaData3}, includes $210$ labeled ceT1 images and $210$ unlabeled hrT2 images. 
Due to the varying voxel spacing in the training data, we resampled the images into a common spacing of $0.6 \times 0.6 \times 1.0$ mm and normalized the intensity to [0, 1] using the Min-Max scaling. 
To remove the noise, the images were cropped into $256\times256$ pixels in the xy-plane using the $75$ percentile binary threshold~\cite{choi2022using}, resulting in $256 \times 256 
\times$ N image volumes for 3D nnU-Net training. The processed 3D volumes were sliced along the z-axis to produce N number of 2D image training samples for the 2D CUT. 
The Dice Score (DSC [\%])~\cite{sudre2017generalised} and the Average Symmetric Surface Distance(ASSD [voxel])~\cite{lu2017automatic} were used to assess the model performance on VS and Cochlea segmentation.




\subsection{Implementation details}

We used single NVIDIA A40 GPU with 48GB of memory for model training.
We followed~\cite{park2020cut} to optimize the proposed segmentation-enhanced CUT network, in which the weights for adversarial loss, contrastive loss, and segmentation loss were set the same. 
Following~\cite{chen2020unsupervised}, the loss weights for additional segmentation in the modified CUT were set to $1$ and $0.1$ for the last layer and the second last downsampling layers respectively.
For pseudo-labeling, we closely followed the nnU-Net optimized settings~\cite{nnunet} and trained for $200$ epochs with the generated hrT2 images and augmented hrT2 images using cross-validation.
The segmentation results from an ensemble of five-fold cross-validations on unlabeled hrT2 images were then used as pseudo labels for the following processes.
For the multi-scale mean teacher, the EMA update $\alpha$ was set to $0.9$, and the loss weights for consistency regularization were set to  $\{0.05, 0.05, 0.05, 0.4, 0.5\}$, assigned to each feature map according to its size in ascending order, and we followed the deep supervision scheme in nnU-Net~\cite{nnunet}.
The loss weights were ramped up across $160$ epochs using the sigmoid function.
We used a combination of Dice and cross-entropy losses as our objective function and trained the MS-MT model with an initial learning rate of $0.01$ using stochastic gradient descent for $300$ epochs. 
Following~\cite{meanteacher}, the results from the five-fold teacher models were ensembled, and then post-processed by computing the largest connected component (LCC) to remove unconnected labels (The first LCC was preserved for the VS while the first and second LCCs were preserved for the Cochlea) for final submission.

\begin{table}[t]
\centering
\setlength\tabcolsep{7pt}
\caption{Quantitative results on the validation dataset. The metrics are presented in the format of mean±std.}
\label{tab:results}
\scalebox{0.9}{
\begin{tabular}{|c|c|c|c|c|}
\hline
VS Dice         & VS ASSD         & Cochlea Dice    & Cochlea ASSD    & Mean Dice      \\ \hline
0.8380±0.097 & 0.5555±0.3647 & 0.8139±0.0312 & 0.2652±0.1553 & 0.8260±0.0508 \\ \hline
\end{tabular}
}
\end{table}

\subsection{Experimental results}

\begin{figure}[h]
    \centering
    \includegraphics[width = 0.9\columnwidth]{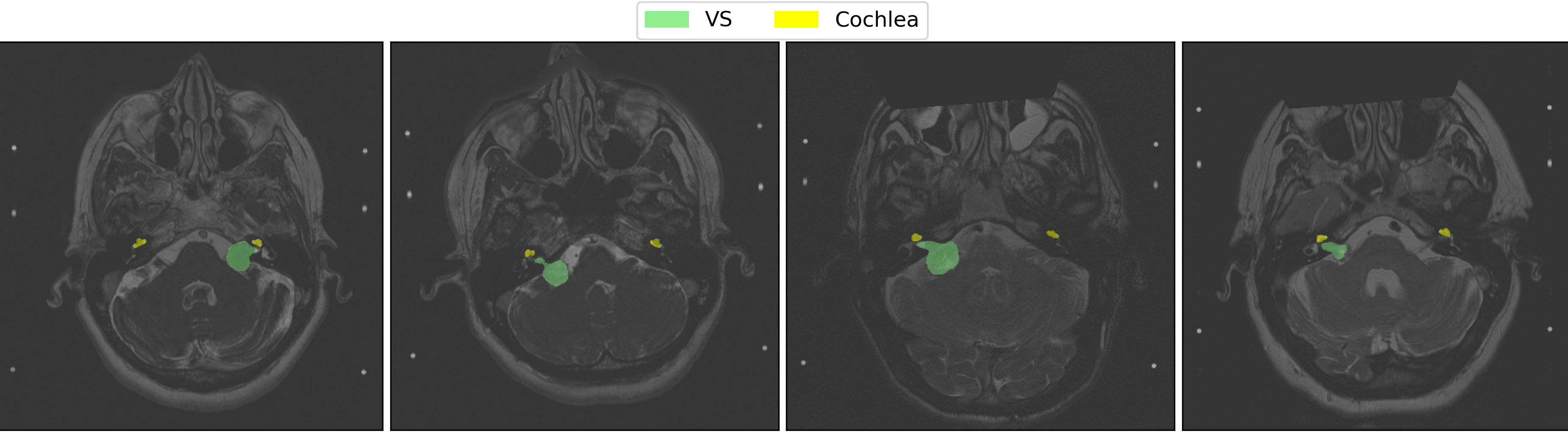}
    \caption{Qualitative results produced by our method. The VS and Cochlea are indicated in \textcolor{green}{green} and \textcolor{yellow}{yellow} color, respectively.}
    \label{fig:sample}
\end{figure}

The results of our method on the validation dataset are presented in Table.~\ref{tab:results}. 
With a mean Dice score of 0.826, our method secures a top-10 finish in the CrossMoDA 2022 competition.
Fig.~\ref{fig:sample} highlights some qualitative results produced by our method on the validation set. Finally, our method ranked 5th in the test phase of the crossMoDA 2022 challenge with a mean Dice score of 85.65\%, and ranked in 3rd place for the Vestibular Schwannoma segmentation with a mean
Dice score of 86.7\%.

\subsection{Ablation study}
Table~\ref{tab:ablation} shows the results of methods of different components,~\emph{e.g.,} intensity augmentation(IA), pseudo-labeling (PL), and mean teacher(MT) on the validation dataset. We start by training a 3D nnU-Net on generated images by CUT,~\emph{i.e.,} nnU-Net+CUT, which achieves a mean dice of 78.95\%. By adding augmented images into the training process, we can observe an improvement in the segmentation performance of both Cochlea and VS. Then, we introduce pseudo-labeling into the training process, which leads to 1.06\% increments in mean Dice, and a continued increase was observed with the inclusion of self-ensembling learning. We further build a multi-scale mean teacher, which achieved better performance. Finally, compared to CUT, the model trained on generated images with the proposed segmentation-enhanced CUT (SE-CUT) can achieve better performance.

\begin{table}[ht]
\centering
\setlength\tabcolsep{7pt}
\caption{Quantitative results (Dice Score [\%]) by different methods on the validation dataset.}
\label{tab:ablation}
\scalebox{0.9}{
\begin{tabular}{lccc} 
\toprule
Method                                           & Cochlea         & VS              & Mean             \\ 
\hline
nnU-Net+CUT                                 & 79.67±3.72 & 78.22±21.04 & 78.95±11.13  \\
nnU-Net+CUT+IA                            & 80.48±4.14 & 78.31±21.00 & 79.39±11.28  \\
nnU-Net+CUT+IA+PL                       & 81.06±3.42 & 79.85±12.88 & 80.45±6.60  \\
nnU-Net+CUT+IA+PL+MT                  & 81.06±3.28 & 82.67±11.89 & 81.86±6.18  \\
nnU-Net+CUT+IA+PL+MS-MT               & 81.24±3.31 & 82.98±10.25 & 82.11±5.39  \\
nnU-Net+SE-CUT+IA+PL+MS-MT & 81.39±3.12 & 83.80±9.72 & 82.60±5.08  \\
\bottomrule
\end{tabular}
}
\end{table}

\section{Discussion}
In this work, we propose to explore image-level domain adaptation and semi-supervised learning to address the unsupervised domain adaptation problem in cross-modality vestibular schwannoma (VS) and cochlea segmentation. Additionally, intensity augmentation is performed to generate additional training data with different VS and cochlea intensities for diversifying the data distributions. Our final submission achieved a mean Dice score of 82.6\% and 85.65\% in the validation phase and the testing phase, respectively. In particular, our method achieved 5th place for overall segmentation performance and ranked 3rd for Vestibular Schwannoma segmentation. From our experiments, we observe that image adaptation contributes the most to the adaptation performance. We can regard the 3D nnU-Net trained on generated target images with CUT (\emph{i.e.,} nnU-Net+CUT) as a strong baseline, achieving dice scores of approximately 80\%, which signifies a competitive result within the leaderboards. However, synthetic images may include noises and artifacts, which would influence the follow-up segmentation performance. Many works~\cite{shin2022cosmos,park2020cut} have been proposed to address the limitations of GANs. In our work, we design a segmentation-enhanced translation network for improved adaptation performance. For data augmentation, we adjust the intensities of VS and cochlea based on clinical characteristics and medical knowledge and achieved better performance. Besides, spatial augmentation methods~\cite{zhao2022meta} can be further explored in this UDA task. For SSL, we adopted pseudo-labeling and self-ensembling learning methodologies to improve the adaptation performance. According to our experimental results, both SSL methods are beneficial to improved segmentation performance. However, one limitation of pseudo-labeling is that low-quality pseudo-labels would adversely influence the training process. Since different perturbations are introduced to the self-ensembling models, the influence of noise from synthetic images and pseudo-labels could be weakened to some extent. We further advance multi-level consistency and deep supervision into self-ensembling learning, achieving better UDA performance.

\section{Conclusion}
In this work, we proposed a three-stage UDA method based on image translation, pseudo-labeling and multi-scale self-ensembling learning to close the gap between two domains and conduct effective image segmentation on the target domain. Specifically, we worked with annotated ceT1 and unannotated hrT2 data to perform segmentation on the target hrT2 domain. In addition, intensity augmentation was implemented to improve the generalizability of the model. With a mean Dice of 0.8216, we achieved a top 10 finish in the CrossMoDA 2022 validation phase and eventually, placed 5th overall in the final competition ranking, demonstrating its effectiveness in bridging the gap between the ceT1 and hrT2 MRI modalities.

%
%
%
\bibliographystyle{splncs04}
\bibliography{refs}

\end{document}